\documentclass[10pt,preprint]{aastex}

\newcommand{\etal}{{\frenchspacing\it et al. }}

\newcommand{\lsim}{\hbox{ \rlap{\raise 0.425ex\hbox{$<$}}\lower 0.65ex\hbox{$\sim$} }}
\newcommand{\gsim}{\hbox{ \rlap{\raise 0.425ex\hbox{$>$}}\lower 0.65ex\hbox{$\sim$} }}

\shorttitle{Palomar Abell Cluster Optical Survey I.}
\shortauthors{Gal \etal}

\begin{document}
\singlespace

\title{The Palomar Abell Cluster Optical Survey I: \\
  Photometric Redshifts for 431 Abell Clusters}

\author{R.R. Gal\altaffilmark{1}, R.R. de Carvalho\altaffilmark{2}, R. Brunner, S.C. Odewahn\altaffilmark{3} \& S.G. Djorgovski}
\affil{Palomar Observatory, Caltech, MC105-24, Pasadena, CA 91125}

\altaffiltext{1}{Email: rrg@astro.caltech.edu}
\altaffiltext{2}{Observat\'orio Nacional, Rua Gal.  Jos\'e Cristino,
77 -- 20921-400, Rio de Janeiro, RJ, Brazil}
\altaffiltext{3}{Arizona State University, Dept. of Physics \& Astronomy,
Tempe, AZ 85287} 

\begin{abstract}
This paper presents photometric redshifts for 431 Abell clusters imaged as part of the Palomar Abell Cluster Optical Survey (PACOS), of which 236 are new redshifts. We have obtained moderately deep, 3--band (Gunn $gri$) imaging for this sample at the Palomar Observatory $60''$ telescope, as part of the photometric calibration of DPOSS. Our data acquisition, reduction, and photometric calibration techniques are described, and photometric accuracy and consistency is demonstrated. An empirical redshift estimator is presented, utilizing background-corrected median $g-r$ colors and mean $g$ magnitudes for the ensemble of galaxies in each field. We present photometric redshift estimates for the clusters in our sample with an accuracy of $\sigma_z=0.038$. These redshift estimates provide checks on single-galaxy cluster redshifts, as well as distance information for studies of the Butcher-Oemler effect, luminosity functions, $M/L$ ratios, and many other projects.

\end{abstract}


\section{Introduction}
Clusters of galaxies are the largest bound systems in the Universe providing useful constraints for theories of 
large-scale structure formation and evolution. They are the samples of choice for studying galaxy evolution in dense environments, with many tens or even hundreds of galaxies in a small, physically associated volume. Multicolor optical photometry of galaxy clusters is commonly used to study the Butcher-Oemler effect (Butcher \& Oemler 1978), the morphology-density relation (Dressler 1980), and other correlations between overall cluster properties, galaxy properties, and redshift. Comparisons between optical and X--ray properties of galaxy clusters are also of considerable scientific interest. For instance, mass-to-light ratios of clusters are also useful for constraining cosmological parameters, including $\Omega$, the mass density of the universe. Properly understood catalogs in the optical and X--ray can help us better understand the various selection effects present in both types of cluster samples.

To obtain the maximal scientific return from such studies, it is necessary to know the redshifts of the clusters. Unfortunately, the majority of known galaxy clusters do not have measured redshifts. Nearly fifty years has passed since the publication of Abell's (1958) optically selected cluster catalog, and only $\sim 1/3$ of the Northern clusters have had spectroscopically measured redshifts, with many of these based on only one or two galaxies. Even at low redshift, obtaining accurate cluster redshifts requires a four-meter class telescope with multi-object spectroscopic capability; performing a survey of hundreds or even thousands of clusters is prohibitively time consuming.

In recent years, there has been an increasing recognition that
redshifts of individual objects or clusters can be estimated quite
accurately from photometric data (Frei \& Gunn 1994, Brunner \etal
1997, {\it etc.})  These estimators have traditionally relied on
either empirical correlations between individual galaxy colors and
spectroscopically measured redshifts (Connolly \etal 1995 ), or a
template method wherein model spectra are created from evolutionary synthesis codes or
spectroscopic data (Gwyn \& Hartwick 1996). However, both techniques aim to measure the redshifts of single
objects using many ($n\ge4$) colors; increasing the number of colors
results in more accurate redshifts over a larger redshift range. An
extensive discussion and comparison of existing techniques can be
found in Hogg \etal (1998).  Unlike these methods, we
are instead relying on only two filters (Gunn $g$ and $r$) to derive
photometric redshifts for an ensemble of objects (a galaxy cluster)
over a relatively small redshift range ($0<z<0.3$). Prior methods for
estimating cluster redshifts have generally relied on the magnitude of
the $n$-th brightest galaxy (Abell 1958, Dalton \etal 1994).
 
Using a small (1-meter class) telescope, equipped with a large-format CCD, one can obtain photometric data on a very large sample of clusters. Greater amounts of observing time are more readily scheduled on such telescopes, and the integration times required for imaging are much shorter than for spectroscopy to comparable depths on larger telescopes. In addition, many optical imaging surveys, such as the Digitized Second Palomar Sky Survey (DPOSS, Djorgovski \etal 1999) and the Sloan Digital Sky Survey (SDSS, Gunn \& Weinberg 1995), provide photometric data which can be used for this purpose. For instance, Gal \etal (2000) presented a simple photometric estimator for galaxy clusters found in DPOSS, and the SDSS photometric system (Fukugita \etal 1996) was designed specifically to improve photometric redshift estimation. 

The success of these estimators spurred us to utilize our data on Abell clusters to measure their redshifts photometrically. We have observed 468 Abell clusters at the Palomar $60''$ telescope, using two different detectors, in the Gunn $gri$ filters (Thuan \& Gunn 1976, Wade \etal 1979). The primary purpose of these data are to provide precision photometric calibration for DPOSS. Nevertheless, this large, homogeneous dataset is a valuable resource in and of itself, and is ideal for photometric redshift estimation.

In the first section of this paper, we describe the telescope, CCD cameras, and data taking strategy used during the course of this survey. The second section describes the observing program. The third section presents our data reduction procedure and the derivation of photometric calibration, using Gunn standards which were imaged every night. We also demonstrate our photometric accuracy using data acquired on the same clusters on multiple nights and with different detectors. The fourth section
presents the photometric redshift estimation technique, and our estimated redshift errors, as well as a table of the measured redshifts. We conclude with a brief discussion of our results.

\section{Data Acquisition and Analysis}

\subsection{Telescope and Detectors}

All data described in this paper were obtained at the Palomar Observatory $60''$ telescope. The telescope is an $f/8.75$ Ritchey-Chretien design, with the CCD imaging cameras placed at Cassegrain focus. From the initial data through June 1996, and on rare occasions thereafter, we used CCD16, a SITe $1024^2$, thinned,
 AR coated array, with $24\mu m$ pixels. The pixel scale is $0.376''$/pixel, providing a $6.4'\times6.4'$ field of view. CCD16 has a gain of $2.5e^-/DN$ and a read noise of
 $8.2 e^-$. Starting in July 1996, a new, larger detector was made available at the Palomar 60''. This detector, CCD13, is a SITe $2048^2$ thinned, AR coated array, also with $24\mu m$ pixels. The pixel scale at this detector is $0.368''$/pixel, providing a $12.56'\times12.56'$ FOV. CCD13 has a gain of $1.63e^-/DN$ and a read noise of $6.3e^-$. In addition to the factor of four increase
 in area over CCD16, this detector also provides extremely good blue sensitivity ($\sim55\%\ QED$ at 4000\AA). All objects were observed in the Gunn $gri$ filters. 

\subsection{Observations}

The imaging targets are selected from the list of northern Abell Clusters (Abell, Corwin \& Olowin 1989). Because the survey was designed to provide photometric calibration and star-galaxy separation data for DPOSS (Djorgovski \etal 1999), we attempted to observe at least two Abell clusters per DPOSS field (each field being $\sim36$ square degrees). Priority was given to the richest clusters closest to the plate centers; therefore, our sample is biased toward richer Abell clusters. Nevertheless, many plates have only one or two known clusters; in such cases, whatever clusters were available were observed. Approximately fifty clusters were observed twice, sometimes with both detectors. This allows us to check our photometric accuracy, as well as consistency between detectors.

Data were taken only on photometric nights with seeing better than $2''$. The mean seeing for our data is $\sim1.5''$; the best seeing is $0.9''$.
Integration times were fixed for each filter/CCD combination, regardless of sky brightness or seeing. The vast majority of nights allocated for this program were $>75\%$ dark. For CCD16, we integrated for $1200s$ in $g$, and $900s$ each in $r$ and $i$. When using CCD13, the integration times were shortened to $900s$ in $g$, and $600s$ each in $r$ and $i$. These yielded limiting magnitudes of $m_{lim,gri}\sim22^m$ for CCD13 observations, with limiting magnitudes for CCD16 observations $\sim0.5$ magnitudes brighter. With the typical observational overhead, we usually observed between seven and eleven clusters in a single night. 

For every night that was deemed photometric, we observed a set of Gunn standards. Between 5 and 12 observations of available Gunn standards were made each night, at a variety of airmasses. Each star observed at each pointing was observed three times on a single frame by closing the shutter, offsetting the telescope, and reopening the shutter. This was done to increase the S/N of our photometric measurements, as well as avoid excessive time loss due to the long ($\sim180$s) readout time for CCD13. In addition to the standard stars, we also took nine bias frames, nine dome flats in each filter, and three sky flats in each filter on every night of observing.

\section{Data Processing}

\subsection{Image Reduction}
Data were processed using the IRAF data reduction package (Tody 1986). First, a single bias value, determined from the median value in the overscan region of each image, was subtracted from all object and calibration frames. The bias frames were median filtered and fit with a smooth polynomial, which was subtracted from all object and flat field images. Individual dome and sky flats in each filter were  median stacked, and the resulting sky flats processed with the dome flats. The remaining large scale structure was smoothed, and this smooth correction was recombined with the dome flats to create a master flat field image for each filter.

These master flat field corrections were applied to all target and standard star frames. After this procedure there were still noticeable large scale gradients in the CCD illumination pattern, especially in the $g$ frames. To remove this, we generated dark sky flat field frames (also known as illumination corrections) by median filtering all of the unregistered target frames in each filter. On those nights where too few targets were observed to generate such an illumination correction, data were combined with an adjacent night. These combined images were smoothed with a large (50 pixel) boxcar, and the resulting correction applied to both the target and standard star frames.  This procedure results in images that are flat to the $\sim1\%$ level. Finally, for CCD13, it was necessary to apply a fringe correction to the $i$ images. This was generated by median stacking the completely flat-fielded $i$-band target images, and smoothing the result with a small (5 pixel) boxcar. The resulting image was used as the fringe correction for the target frames.

\subsection{Photometric Calibration}
Standard star photometry was performed using the {\it apphot} package in IRAF. Stars were photometered in numerous apertures up to 50 pixels ($18.5''$) in radius, and the local sky was determined using a 10 pixel wide annulus centered on each star, starting at a radius of 50 pixels. The convergence magnitudes for all three exposures of each star on each frame were measured, and the three values averaged to provide a mean instrumental magnitude.

The resulting collection of between five and twelve measured instrumental magnitudes were used to determine the zero-point offset, airmass term, and color term in each filter. We used the IRAF {\it fitparams} task to fit the relation
\begin{equation}
{m_{true} = m_{inst} + A + Bsec(z) + C(color)}
\end{equation}
for each filter. On some nights, not enough standards were observed to robustly determine values for all three unknowns. In these cases, the airmass and color terms were fixed at those derived from the mean values from nights with sufficient standards (given in Table 1), and only the zero point was determined. Typical {\it rms} deviations in the fit of the calibration relation are 0.01 magnitude. In addition, on two nights, over twenty Gunn standards were observed to test our measurements of the airmass and color terms. In both cases, the derived terms were within $10\%$ of the mean values derived from other nights with many fewer standards.

\begin{deluxetable}{clr}
\tablenum{1}
\tablecolumns{3}  
\tablewidth{0pt}  
\tablecaption{CCD13 Extinction and Color Terms}
\tablehead{\colhead{Filter} & \colhead{Airmass} & \colhead{Color}}
\startdata

$g$ & $-0.152$ & $0.150\times(g-r)$  \\
$r$ & $-0.094$ & $0.068\times(g-r)$ \\
$i$ & $-0.070$ & $-0.013\times(g-i)$  \\
\enddata
\end{deluxetable}

\subsection{Object Detection and Photometry}

Object detection on the target frames was done using the FOCAS package (Jarvis \& Tyson 1979, Valdes 1982). The $g$, $r$, and $i$ frames were processed independently, using detection parameters of $2.5\sigma$ per pixel, a 25 pixel minimum area, and a sky value estimated individually for each image. Object classification was also performed by FOCAS, and the classifications were visually inspected. Bright objects with incorrect classifications (usually due to saturated pixels) were corrected by hand. The photometric coefficients derived from the standard star observations were used to determine object magnitudes. The color terms were applied only after objects are detected in multiple filters, and matched together at a later stage. For objects detected in only one filter, default colors of $g-r=0.5$  and $g-i=0.5$ were used when applying the color correction. 
In order to properly study such a large collection of imaging data, we must establish photometric consistency between many observing nights, and in our case, two different detectors. Photometric accuracy is also a necessary condition for using this data to calibrate DPOSS. To examine this question, we observed 51 clusters on two or more occasions. In some cases, both sets of observations were taken with the same CCD, and in others, with both CCDs. 

We have compared the photometry of numerous, multiply-observed clusters in three scenarios: {\it(1)} both observations with CCD16, {\it(2)} one observation with CCD16 and one with CCD13, and {\it(3)} both observations with CCD13. We find that there are no systematic night-to-night or CCD-to-CCD photometric offsets greater than $0.05^m$, and the residuals are typically $0.05^m$ at $m_{gri}=19^m$, rising to no more than $0.15^m$ for objects within one magnitude of the detection limit. Table 2 presents the typical residuals at $m_r=20.0^m$ for the various comparisons, while Figure 1 shows a typical set of photometric comparisons. The left plot is a comparison of photometry for the cluster A31, where both nights of data were taken with CCD16. The data were taken 14 months apart. The center plot compares photometry for the cluster A98, with one night of CCD16 data, and one night of CCD13 data. In this case, the data were taken almost four years apart. Finally, the right plot shows photometry for A2562, with both nights of data taken with CCD13, taken only 3 months apart. The bottom panels show the magnitude differences between each pair of observations.

\begin{figure}
%\epsscale{0.5}
\plotone{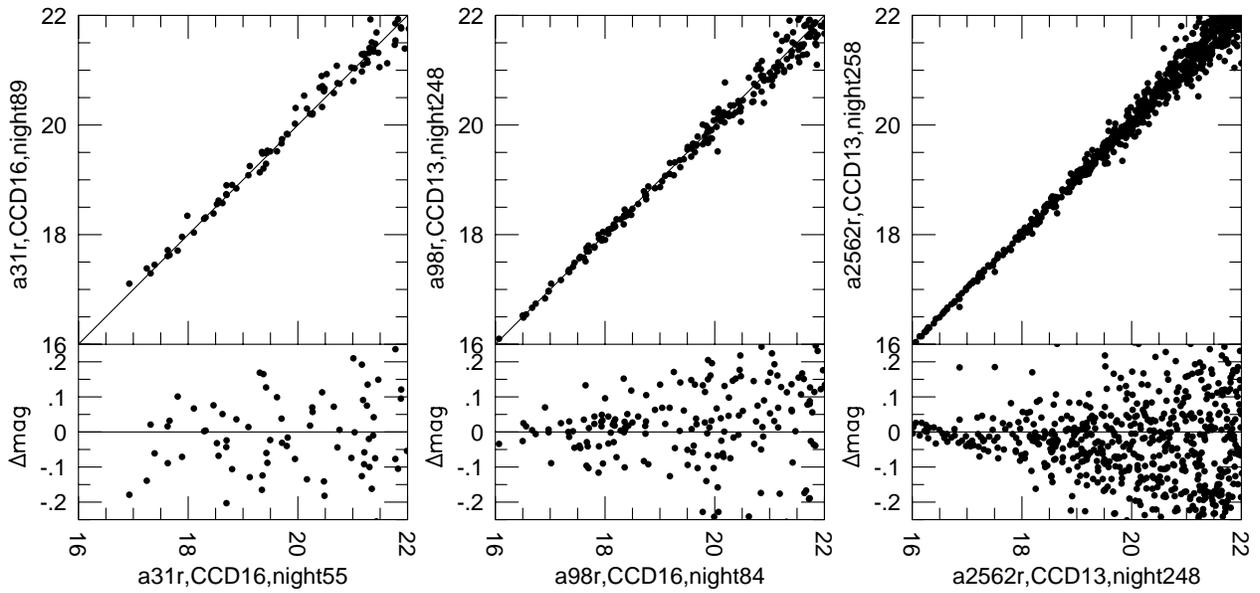}
\caption{Typical photometric comparisons. The left plot is a comparison of photometry for the cluster A31, where both nights of data were taken with CCD16. The center plot compares photometry for the cluster A98, with one night of CCD16 data, and one night of CCD13 data. Finally, the right plot shows photometry for A2562, with both nights of data taken with CCD13. The bottom panels show the magnitude differences between each pair of observations.}
\end{figure}

\begin{deluxetable}{llll}
\tablenum{2}
\tablecolumns{4}  
\tablewidth{0pt}  
\tablecaption{Photometric Residuals at $m_r=20.0^m$}
\tablehead{\colhead{Comparison} & \colhead{$\sigma_g$} & \colhead{$\sigma_r$} & \colhead{$\sigma_i$}}
\startdata
CCD16 {\it vs.} CCD16 & 0.19 & 0.13 & 0.09 \\
CCD13 {\it vs.} CCD16 & 0.10 & 0.09 & 0.10 \\
CCD13 {\it vs.} CCD13 & 0.12 & 0.10 & 0.16 \\
\enddata
\end{deluxetable}

\section{Photometric Redshifts}

From the CCD data, we wish to measure the redshift of each cluster. This can be done either by estimating the redshift of each galaxy individually, and examining the resulting redshift distribution, or by using the  properties of all the galaxies in the image. We have chosen to use the second method, for a number of reasons. First, we only have limited wavelength coverage. Having only three filters (and especially the lack of $u$ images), makes it difficult to disentangle color-redshift degeneracies. Second, our photometric errors on individual objects are significant enough to pose a problem for an object-by-object redshift estimator. Finally, since we have performed object detection in the different filters independently (i.e. we do not have matched apertures), there may be systematic effects in the relative colors of different galaxy types. We have therefore elected to use the average properties of the galaxies in each field. Spectroscopic redshifts are taken from Struble \& Rood (1991,1999).

\subsection{Deriving the Relation}
We estimate the redshift assuming that each field contains a single
cluster, at one redshift, and the cluster galaxy population is
dominated by early-type galaxies.  The $g-r$ color of elliptical
galaxies evolves rapidly at $z<0.4$ due to a strong $k$-correction, as
the 4000\AA \space break passes through the $g$ filter. Similarly, the
$g$ magnitudes of these galaxies fade rapidly with redshift, due to
both distance and $k$-correction effects. For these reasons, we have
chosen to use the $g-r$ colors and $g$ magnitudes of the cluster
galaxies in our photometric redshift estimator. The asssumption of a
single cluster per CCD field is reasonable, although projections are
certain to occur. While it has been estimated that up
to 35\% of clusters in the Abell catalog with $R\ge1$ may be the result
of projections of poorer clusters  (van Kaarlem, Frenk \& White 1997),
our results will be dominated by the richest cluster in each field. Our
redshift estimator will fail if there are two similarly rich clusters 
projected along the line of sight.

We count the number of galaxies as a function of color, $N_{g-r}$, and the number as a function of $g$ aperture magnitude (where our aperture has a radius of 13 pixels, or $\sim5''$), $N_g$, inside each CCD frame, imposing a magnitude limit of $m_r=21.5^m$ for CCD 13 and $m_r=21.0^m$ for CCD16, to avoid incompleteness. We use the whole CCD area (as opposed to a given physical radius), because the size of our field is comparable to the sizes of clusters in the redshift range we are examining, and the centers of many clusters are not well determined. 

First, we apply an extinction correction to the magnitudes and colors of all objects in our fields, derived from the maps of Schlegel \etal (1998). A single correction is used for each CCD field, since the pixel size of the extinction maps is comparable to the CCD FOV, and the vast majority of our fields are at high galactic latitudes; the mean $E(B-V)$ for our sample is $0.075^m$. A statistical background correction must then be applied to both galaxy color and magnitude distributions. These corrections, $N_{bg,g-r}$ and $N_{bg,g}$ are determined from a set of 22 observations of random fields taken with CCD13, during numerous observing runs.  This distribution (scaled to the appropriate area for CCD16 data), is then subtracted from the color and magnitude distributions of each cluster, and the median $g-r$ color and mean $g$ magnitude of the remaining galaxies calculated. An example showing this procedure is shown in Figure 2. The thin solid line is the distribution of all galaxy colors ($g-r$) in CCD13 images of Abell 2063. The dotted line is the distribution of galaxy colors derived from our random field observations, while the thick solid line is the background-corrected distribution. A sharp peak at $(g-r)\sim0.5$ is seen, corresponding to the early-type galaxy population in this cluster.

\begin{figure}
%\epsscale{0.5}
\plotone{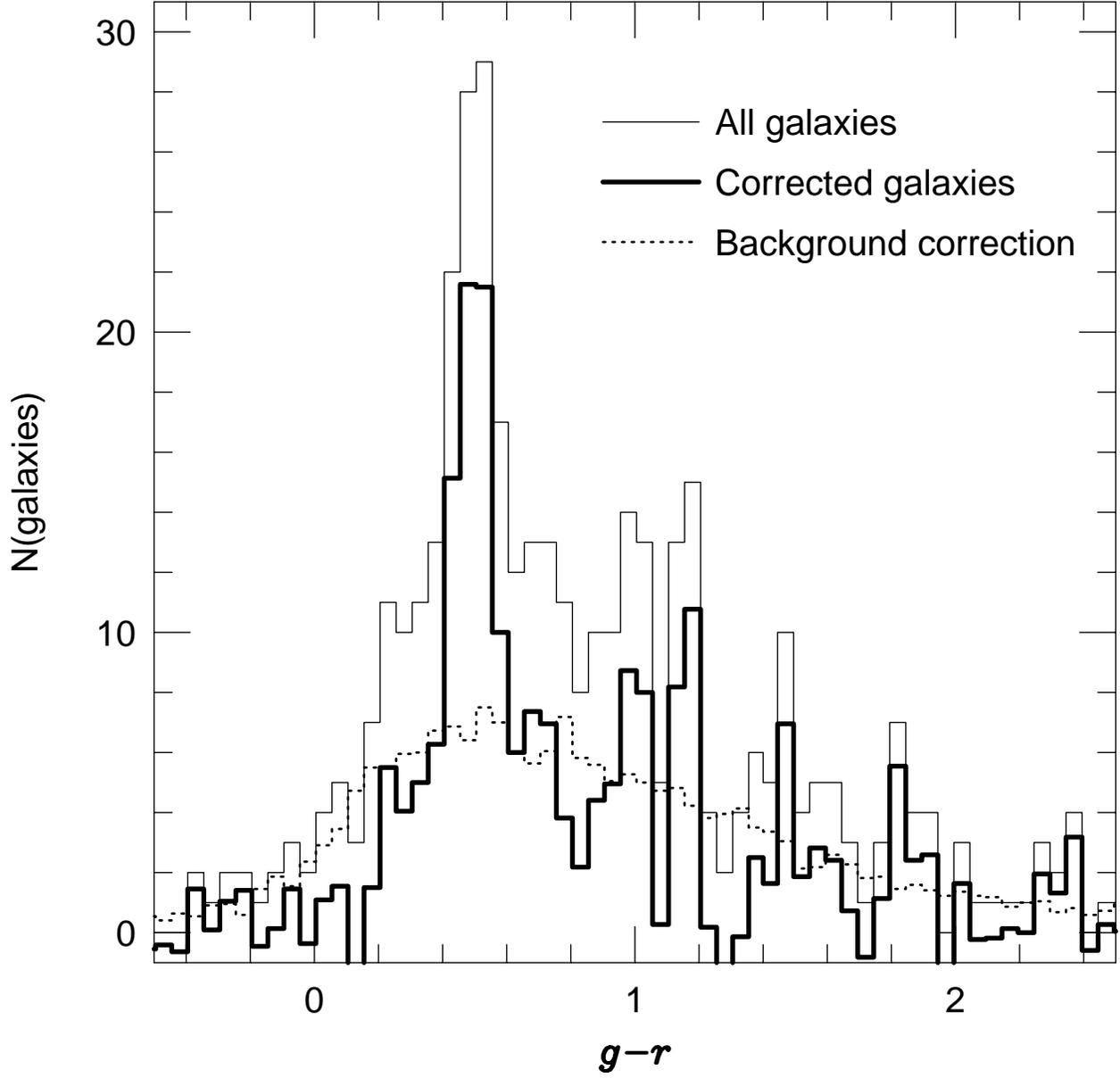}
\caption{An example of the background correction procedure. The thin solid line is the distribution of all galaxy colors ($g-r$) in CCD13 images of Abell 2063. The dotted line is the distribution of galaxy colors derived from our random field observations, while the thick solid line is the background-corrected distribution. A sharp peak at $(g-r)\sim0.5$ is seen, corresponding to the early-type galaxy population in this cluster. }
\end{figure}

Empirical relations between our measured cluster properties and the spectroscopic redshifts were then derived independently for CCD13 and CCD16 data.  For CCD13, 114 clusters with measured redshifts were used, while only 33 clusters were available for CCD16. Using the GAUSSFIT package (Jefferys \etal 1988), we performed a bivariate least-squares fit, deriving the following relations between redshift, median $g-r$ color, and mean $g$ magnitude:

{\noindent \bf CCD13, with $m_{r,lim}=21.5$:}
\begin{equation}
{log(z) = 3.2619 \times (g-r)_{med} - 1.6687 \times (g-r)_{med}^2 + 0.1190 \times g_{mean} - 4.6532}
\end{equation}
{\noindent \bf CCD16, with $m_{r,lim}=21.0$:}
\begin{equation}
{log(z) = 3.2151 \times (g-r)_{med} - 1.6957 \times (g-r)_{med}^2 + 0.0710 \times g_{mean} - 3.5983}
\end{equation}
 
The formal errors on each coefficient from the fit are very small ($\pm0.0002$ or less).  The above relations, although apparently dissimilar, should be universal, since we are using calibrated quantities. To test this, we applied the same magnitude cut ($m_r<21.0$) used for the CCD16 data to the CCD13 data.  The coefficients from the CCD16 fit were then used to estimate redshifts for the CCD13 clusters with spectroscopic redshifts. The $rms$ of $z_{spect} - z_{phot}$ was only marginally higher (0.025 vs. 0.024) than using the true CCD13 fit, although a slight systematic overestimate of the redshift does occur at $z<0.1$. The effects of detector area, the background correction technique, and the sample used can all affect the fit; nevertheless, it appears that our derived relations should be applicable to other data sets treated in an identical fashion ({\it i.e.} same filters, apertures, magnitude cuts, and background correction).

It is important to note that the combination of both color and magnitude information significantly improves our photometric redshift estimates. Figure 3 shows the photometric redshift estimates using magnitudes only, colors only, and the combination of colors and magnitudes for the CCD13 data. The scatter in redshift decreases from $0.034$ for magnitudes only, to $0.031$ for colors only, to $0.024$ when both are used.

\begin{figure}
%\epsscale{0.5}
\plotone{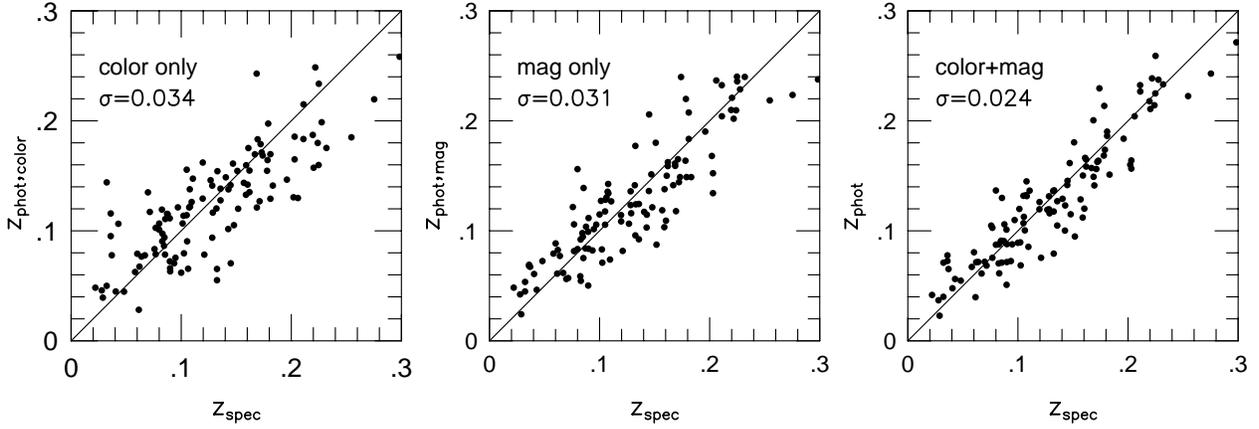}
\caption{The photometric redshift estimates for the CCD13 data plotted against the spectroscopic redshifts, using magnitudes only, colors only, and the combination of colors and magnitudes to derive the estimates. The decrease in scatter is clearly seen.}
\end{figure}

\subsection{Redshift Errors}

The $rms$ of $z_{spect} - z_{phot}$ are $ \sigma(z)=0.024$ and $0.027$ for CCD13 and CCD16, respectively. These figures represent the intrinsic scatter of our derived relationship between redshift and photometric properties. The scatter is larger for the CCD16 data due to the small number of clusters with spectroscopic redshifts used in the derivation. The error on any individual cluster redshift depends not only on the scatter of the relation, but also on the number of clusters in the local redshift range used to determine the relation. For instance, at the high redshift end ($z\sim0.3$), there are only four clusters which constrain the relation for the CCD13 data. Therefore, we expect the errors in this redshift range to be larger than at $0.05<z<0.1$, where there are 29 clusters. 

To estimate the magnitude of this effect, we have performed bootstrap simulations of our training samples. We randomly select clusters from our training sample, with replacement, until we have built a new training sample of the same size as the original (114 for CCD13, 33 for CCD16). This new sample is used to derive a new redshift estimator, which is then applied to the original sample. We repeat this procedure 500 times, and calculate the mean redshift estimate, and the standard deviations about this estimate, for each cluster. Figure 4 shows the photometrically estimated redshift against the spectroscopically measured redshift for both CCDs. Error bars on individual redshifts represent the one-sigma redshift limits from the bootstrap procedure. 

\begin{figure}
%\epsscale{0.5}
\plotone{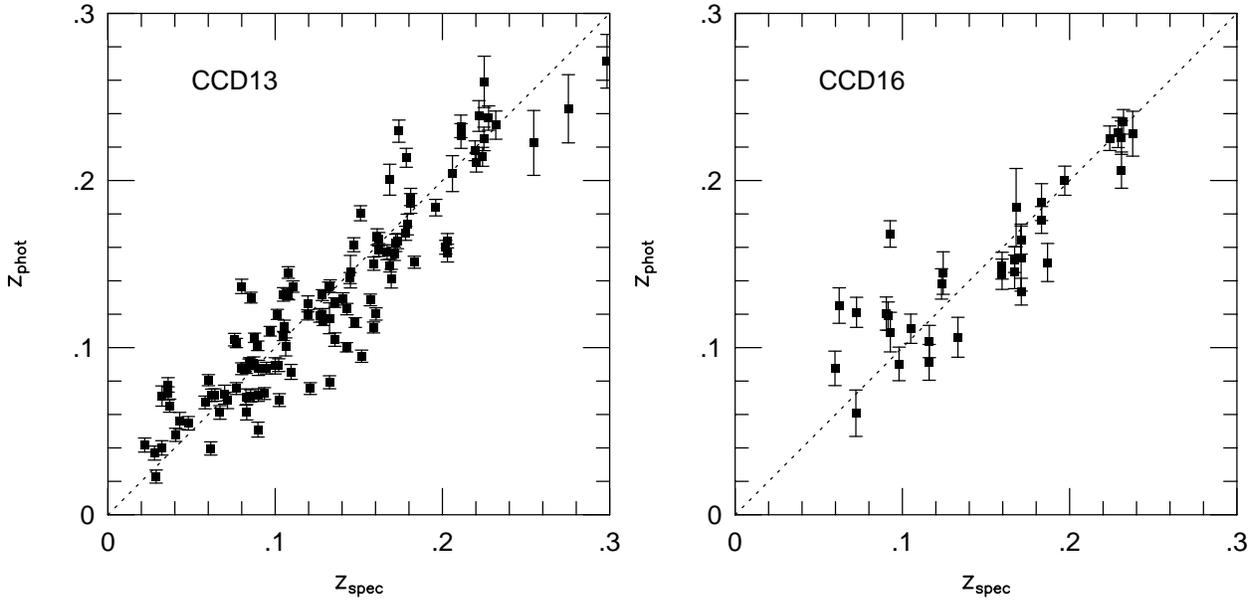}
\caption{The photometrically estimated redshift against the spectroscopically measured redshift for both CCDs. Error bars on individual redshifts represent the one-sigma redshift limits from the bootstrap procedure. The 114 clusters used to derive the relation for CCD13 are on the left, while the 33 clusters used for CCD16 are on the right.}
\end{figure}

Clearly, the errors are smallest in the redshift bins with the largest
number of calibrating clusters. Figure 5 shows the errors on each
cluster redshift derived from the bootstrap procedure. Solid circles
show the mean redshift errors in bins of $\Delta z=0.05$. For CCD13,
the errors are largest at $z\sim0.3$, where there are only 3
clusters. The redshift distribution for CCD16 clusters is more even,
yielding nearly equal errors over the entire redshift range. To
estimate the true redshift error on any given measurement, we add in
quadrature the intrinsic scatter of our $z_{phot}$ vs. $z_{spec}$
relation, and the error from the appropriate bin from our bootstrap
simulation. This error estimate then encompasses both the intrinsic
errors of our fit, and the errors introduced from our sample
selection. For 57 clusters, we have multiple observations from which
to estimate their redshifts. Comparing these redshift estimates, we
find $rms$ $\Delta z=0.038$, which is exactly what we expect from the
individual error estimates. Our redshift errors are not dominated by
photometric errors, since a large number of galaxies contribute to the
mean color and magnitude values in each cluster. The scatter is most
likely due to inaccurate spectroscopic redshifts (less than $20\%$ of
the clusters have three or more galaxies with measured redshifts) and
variance in the cluster properties (blue galaxy fraction, richness,
{\it etc}. For instance, Miller \etal (1999) have demonstrated that
projection effects have resulted in significantly erroneous redshifts
for 14\% of clusters with single spectroscopic redshifts. These two
effects cannot be disentagled without a larger sample of clusters with
properly measured spectroscopic redshifts. Unfortunately, almost all
of the clusters we studied that also have a significant number of
spectroscopic redshifts are very nearby ($z<0.05$), and therefore
we cannot draw any significant conclusions about spectroscopic redshift errors
for our sample as a whole.

W\begin{figure}
%\epsscale{0.5}
\plotone{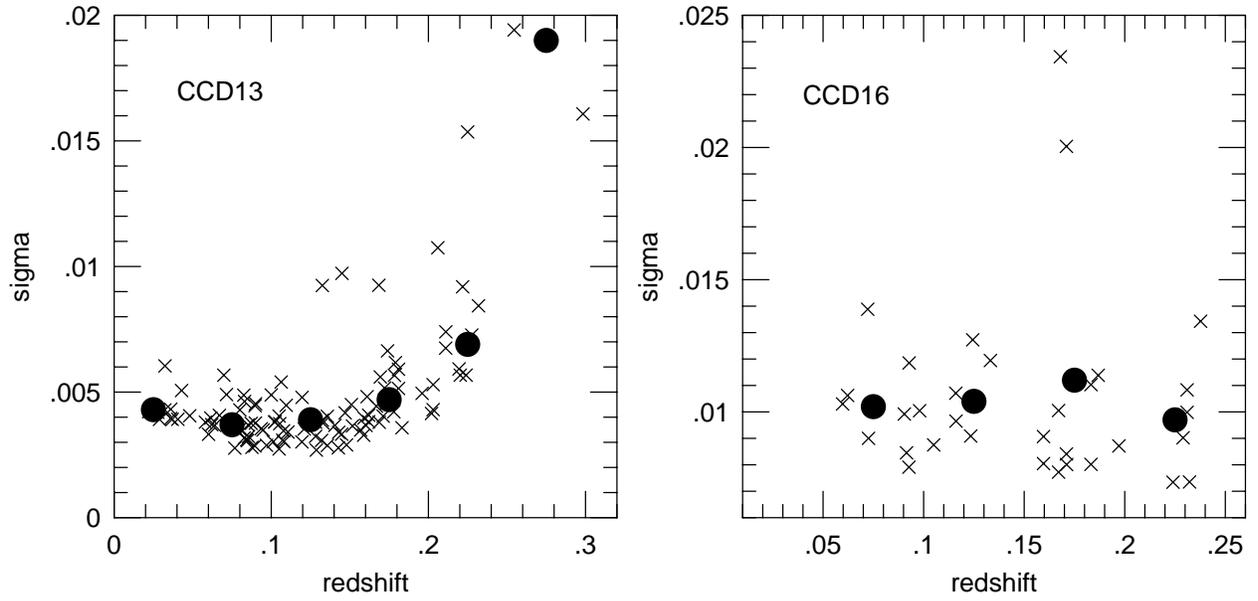}
\caption{Crosses show the error of each cluster redshift, as determined by our bootstrap simulation. Solid circles show the mean redshift errors in redshift bins of $\Delta z=0.05$. The left panel shows the 114 clusters from CCD13, while the right panel shows the 33 clusters from CCD16. }
\end{figure}

e present the photometric results in Table 3. Clusters are arranged in numerical order. The first column gives the Abell cluster number. The second column gives the CCD used for the observation. The next two columns provide the backgorund-corrected median $(g-r)$ color and $g$ magnitude for each cluster. The fifth column is the extinction $E(B-V)$. The sixth column provides the photometric redshift estimate, and the seventh column contains the estimated error, as described above. The final column gives the spectroscopic redshift, if it is available. For clusters with very low spectroscopic redshifts ($z_{spec}<0.03$), we usually obtain a significantly higher photometric redshift. For these clusters, our CCD field covers only a small physical area, and we are likely seeing through the lower redshift cluster and measuring a background cluster.  

\section{Discussion and Future Work}

This paper presents the data acquisition, reduction, and photometric analysis of an unprecedented sample of CCD $gri$ observations of Abell clusters. We have demonstrated the photometric consistency of our sample, and presented a simple yet effective photometric redshift estimator. A total of 431 clusters have been studied, providing 236 new photometric redshift estimates. In addition, we have shown our photometric redshift estimator to be universal. The derived relations between redshift, colors, and magnitudes can be applied to any other data set taken with the same filters and analyzed in an identical way.

These new redshifts enable a variety of projects for which photometry alone is not sufficient, and distance information is required, but high accuracy is not necessary. The evolution of galaxy and cluster properties with time (i.e. the Butcher-Oemler effect, morphology-density relation, luminosity functions, {\it etc.}) are prime examples of such science. In addition, these photometric redshifts are a useful check on the spectroscopic redshifts of galaxy clusters, many of which are derived from a single measured galaxy.

Future papers will make use of this large sample to perform large, statistical studies of low redshift ($z<0.3$) galaxy clusters. The next paper will discuss the Butcher-Oemler effect (as was done for a smaller sample by Margoniner \& de Carvalho 2000), and later papers will measure luminosity functions, study intracluster light, the color-magnitude relation, the Binggeli effect (Binggeli 1982), and galaxy morphology. Bringing in data from other wavelengths, most notably X--ray and radio, we will examine cluster $M/L$ ratios, morphologies, cooling flows, the cluster fundamental plane, and radio  properties of elliptical galaxies. Each paper will focus on a specific study of this vast statistical sample. Finally, the entire CCD data set (images as well as catalogs) will be released to the astronomical community.

\acknowledgments

RRG was supported in part by an NSF Fellowship, NASA GSRP NGT5-50215, and a Kingsley Fellowship. We thank the Norris Foundation for their generous support of the DPOSS project. We also thank the Palomar TAC and Directors for generous time allocations for the DPOSS calibration effort. Numerous past and present Caltech undergraduates (V. Desai, V. Hradecky, J. Meltzer, B. Stalder, J. Hagab, R. Stob, J. Kollmeier) assisted in the taking of the data utilized in this paper.

\clearpage

\renewcommand{\arraystretch}{.6}
\begin{deluxetable}{rccccccc}
\tablenum{3}
%\tablewidth{350pt}
\tablecolumns{8}  
\tablecaption{Cluster Properties \& Photometric Redshifts}
\tablehead{\colhead{Abell} & \colhead{CCD} & \colhead{$(g-r)_{med}$} & \colhead{$g_{mean}$} & \colhead{$E(B-V)$} & \colhead{$z_{phot}$} & \colhead{Error} & \colhead{$z_{spec}$}} 
\startdata
21 & 13 & 0.56984 & 19.1749 & 0.03911 & 0.0881 & 0.0243 & 0.0946 \\
26 & 13 & 0.68976 & 19.8172 & 0.06752 & 0.1448 & 0.0243 & 0.1449 \\
31 & 16 & 0.62472 & 19.9006 & 0.03836 & 0.1452 & 0.0289 & 0.1596 \\
31 & 16 & 0.65092 & 19.7028 & 0.03686 & 0.1498 & 0.0289 & 0.1596 \\
54 & 13 & 0.75757 & 20.4157 & 0.04703 & 0.1947 & 0.0245 & -- \\
59 & 13 & 0.71243 & 20.1403 & 0.05342 & 0.1660 & 0.0245 & -- \\
60 & 13 & 0.97409 & 19.9467 & 0.04778 & 0.2061 & 0.0250 & -- \\
62 & 16 & 0.50123 & 19.4658 & 0.03938 & 0.0933 & 0.0289 & -- \\
65 & 16 & 0.78512 & 20.0763 & 0.06377 & 0.2026 & 0.0287 & 0.1206\tablenotemark{a} \\
68 & 13 & 1.11481 & 20.5031 & 0.09476 & 0.2233 & 0.0250 & 0.2546 \\
69 & 13 & 0.64688 & 20.1062 & 0.05788 & 0.1415 & 0.0243 & 0.1454 \\
69 & 16 & 0.84825 & 19.2846 & 0.05788 & 0.1899 & 0.0292 & 0.1454 \\
73 & 13 & 1.05318 & 20.3368 & 0.09367 & 0.2244 & 0.0250 & -- \\
79 & 16 & 0.69633 & 19.7945 & 0.08488 & 0.1676 & 0.0292 & 0.0927\tablenotemark{a} \\
82 & 13 & 0.64521 & 19.4615 & 0.03514 & 0.1181 & 0.0243 & -- \\
83 & 16 & 0.80527 & 20.2038 & 0.08388 & 0.2119 & 0.0287 & -- \\
98 & 13 & 0.57916 & 19.7885 & 0.04508 & 0.1073 & 0.0243 & 0.1050 \\
98 & 16 & 0.58635 & 19.0310 & 0.04508 & 0.1136 & 0.0289 & 0.1050 \\
110 & 13 & 0.49108 & 19.4000 & 0.04784 & 0.0715 & 0.0243 & -- \\
115 & 16 & 0.90548 & 19.3913 & 0.05764 & 0.1995 & 0.0292 & 0.1971 \\
136 & 16 & 0.90075 & 20.5781 & 0.05666 & 0.2418 & 0.0287 & 0.1569 \\
137 & 16 & 0.58929 & 19.6325 & 0.06651 & 0.1264 & 0.0289 & -- \\
142 & 16 & 1.19339 & 20.3110 & 0.02611 & 0.1846 & 0.0292 & -- \\
143 & 16 & 0.64219 & 20.2976 & 0.07644 & 0.1617 & 0.0292 & -- \\
152 & 13 & 0.51463 & 18.8969 & 0.03561 & 0.0679 & 0.0243 & 0.0581 \\
153 & 13 & 0.61917 & 20.1015 & 0.03076 & 0.1313 & 0.0243 & 0.1279 \\
153 & 13 & 0.64632 & 19.4959 & 0.03076 & 0.1196 & 0.0243 & 0.1279 \\
154 & 13 & 0.50904 & 19.1955 & 0.06283 & 0.0722 & 0.0243 & 0.0624 \\
163 & 16 & 0.62553 & 20.2190 & 0.05280 & 0.1532 & 0.0292 & -- \\
165 & 13 & 0.60361 & 19.7734 & 0.05807 & 0.1149 & 0.0243 & -- \\
167 & 16 & 0.68404 & 20.3669 & 0.05187 & 0.1795 & 0.0292 & -- \\
175 & 13 & 0.47655 & 18.6362 & 0.05604 & 0.0549 & 0.0243 & 0.1292 \\
175 & 13 & 0.60273 & 19.8184 & 0.05604 & 0.1160 & 0.0243 & 0.1292 \\
184 & 13 & 0.90874 & 20.8581 & 0.03488 & 0.2598 & 0.0306 & -- \\
184 & 16 & 0.82114 & 20.6201 & 0.03484 & 0.2306 & 0.0287 & -- \\
191 & 13 & 0.90729 & 20.6865 & 0.06023 & 0.2477 & 0.0250 & -- \\
196 & 13 & 0.91350 & 20.1011 & 0.07062 & 0.2117 & 0.0250 & -- \\
219 & 13 & 0.70417 & 18.9628 & 0.07354 & 0.1182 & 0.0243 & -- \\
227 & 13 & 0.67306 & 20.0422 & 0.07266 & 0.1482 & 0.0243 & 0.1768 \\
234 & 13 & 0.67408 & 20.3887 & 0.06394 & 0.1634 & 0.0245 & 0.1731 \\
245 & 13 & 0.47259 & 19.1110 & 0.05571 & 0.0616 & 0.0243 & 0.0790\tablenotemark{b} \\ 
246 & 13 & 0.53622 & 18.6123 & 0.06930 & 0.0677 & 0.0243 & 0.0700\tablenotemark{b} \\ 
247 & 16 & 0.70622 & 19.9252 & 0.05394 & 0.1745 & 0.0292 & -- \\
249 & 13 & 0.69410 & 20.6258 & 0.07400 & 0.1824 & 0.0245 & -- \\
249 & 13 & 0.85473 & 20.4095 & 0.07359 & 0.2208 & 0.0250 & -- \\
253 & 13 & 1.05273 & 19.1403 & 0.09358 & 0.1617 & 0.0245 & -- \\
257 & 16 & 0.61246 & 19.6717 & 0.05504 & 0.1355 & 0.0289 & 0.0703 \\
258 & 13 & 0.09506 & 18.7597 & 0.10749 & 0.0111 & 0.0244 & -- \\
258 & 13 & 0.15118 & 18.8687 & 0.10749 & 0.0075 & 0.0244 & -- \\
260 & 13 & 0.48142 & 19.2192 & 0.04944 & 0.0656 & 0.0243 & 0.0369 \\
262 & 13 & 0.52546 & 18.2927 & 0.09019 & 0.0598 & 0.0243 & 0.0163 \\
272 & 13 & 0.57489 & 19.7996 & 0.04788 & 0.1062 & 0.0243 & 0.0877 \\
275 & 16 & 0.86527 & 20.2964 & 0.05229 & 0.2268 & 0.0287 & -- \\
278 & 13 & 0.56409 & 19.7509 & 0.05465 & 0.1013 & 0.0243 & 0.0891 \\
288 & 13 & 0.70207 & 19.6760 & 0.06845 & 0.1430 & 0.0243 & -- \\
288 & 13 & 0.71702 & 19.5644 & 0.06845 & 0.1431 & 0.0243 & -- \\
300 & 16 & 0.72066 & 20.2539 & 0.05813 & 0.1890 & 0.0292 & -- \\
307 & 13 & 1.94115 & 19.4678 & 0.09527 & 0.0051 & 0.0244 & -- \\
311 & 13 & 0.49321 & 18.2959 & 0.16897 & 0.0533 & 0.0243 & 0.0649\tablenotemark{b} \\ 
318 & 13 & 0.65820 & 19.8653 & 0.06498 & 0.1363 & 0.0243 & 0.1320 \\
320 & 13 & 0.91160 & 20.1860 & 0.08404 & 0.2165 & 0.0250 & -- \\
330 & 13 & 0.58288 & 20.2481 & 0.09242 & 0.1231 & 0.0243 & -- \\
333 & 16 & 0.70736 & 19.9606 & 0.09204 & 0.1759 & 0.0292 & -- \\
345 & 13 & 0.61722 & 19.2636 & 0.16864 & 0.1038 & 0.0243 & -- \\
347 & 13 & 1.09518 & 17.3148 & 0.05747 & 0.0950 & 0.0243 & 0.0184 \\
349 & 13 & 0.57984 & 19.3553 & 0.06075 & 0.0955 & 0.0243 & -- \\
360 & 13 & 0.85185 & 20.2633 & 0.07576 & 0.2116 & 0.0250 & 0.2205 \\
364 & 13 & 0.66718 & 20.1751 & 0.16586 & 0.1516 & 0.0245 & 0.1800\tablenotemark{b} \\ 
372 & 13 & 0.64544 & 19.8768 & 0.09561 & 0.1324 & 0.0243 & 0.1073 \\
373 & 13 & 0.63530 & 19.4034 & 0.15390 & 0.1486 & 0.0243 & -- \\
373 & 13 & 0.66272 & 20.1405 & 0.15390 & 0.1133 & 0.0243 & -- \\
374 & 13 & 0.61523 & 19.3218 & 0.05121 & 0.1049 & 0.0243 & 0.0757 \\
376 & 13 & 0.49633 & 18.3979 & 0.07008 & 0.0554 & 0.0243 & 0.0481 \\
377 & 13 & 0.74239 & 19.7053 & 0.14500 & 0.1937 & 0.0245 & -- \\
377 & 13 & 0.75779 & 20.3970 & 0.14423 & 0.1560 & 0.0245 & -- \\
382 & 13 & 0.57552 & 19.7137 & 0.05816 & 0.1040 & 0.0243 & -- \\
384 & 16 & 0.97386 & 20.3809 & 0.04618 & 0.2356 & 0.0287 & -- \\
397 & 13 & 0.40766 & 18.5634 & 0.15054 & 0.0405 & 0.0244 & 0.0327 \\
399 & 13 & 0.45026 & 19.8261 & 0.16449 & 0.0686 & 0.0243 & 0.0724 \\
408 & 13 & 0.65978 & 19.6092 & 0.22645 & 0.1275 & 0.0243 & -- \\
410 & 13 & 0.42726 & 19.1095 & 0.10081 & 0.0512 & 0.0243 & 0.0897 \\
411 & 16 & 0.51087 & 19.3963 & 0.08243 & 0.0953 & 0.0289 & 0.1567\tablenotemark{b} \\ 
421 & 13 & 0.50475 & 18.5584 & 0.36592 & 0.0597 & 0.0243 & -- \\
421 & 13 & 0.57261 & 19.1531 & 0.34166 & 0.0884 & 0.0243 & -- \\
426 & 16 & 0.60447 & 19.4184 & 0.17631 & 0.1273 & 0.0289 & 0.0179 \\
427 & 13 & 0.69380 & 19.5394 & 0.19721 & 0.1354 & 0.0243 & -- \\
429 & 13 & 0.66930 & 19.2705 & 0.36639 & 0.1189 & 0.0243 & -- \\
429 & 13 & 0.72971 & 18.8353 & 0.36639 & 0.1201 & 0.0243 & -- \\
436 & 13 & 0.52151 & 18.4605 & 0.27183 & 0.0617 & 0.0243 & -- \\
437 & 13 & 0.55318 & 19.4990 & 0.04673 & 0.0913 & 0.0243 & 0.0847 \\
439 & 13 & 0.63436 & 18.9645 & 0.18273 & 0.1002 & 0.0243 & 0.1068 \\
444 & 13 & 0.56523 & 19.1289 & 0.15907 & 0.0858 & 0.0243 & -- \\
452 & 13 & 0.74202 & 20.1905 & 0.17141 & 0.1781 & 0.0245 & -- \\
461 & 13 & 0.94078 & 19.9688 & 0.09804 & 0.2063 & 0.0250 & -- \\
465 & 13 & 0.65280 & 19.7421 & 0.26737 & 0.1300 & 0.0243 & 0.1300\tablenotemark{b} \\ 
468 & 13 & 0.55633 & 18.9604 & 0.18108 & 0.0796 & 0.0243 & 0.1325 \\
468 & 13 & 0.73595 & 18.7085 & 0.18108 & 0.1174 & 0.0243 & 0.1325 \\
477 & 13 & 0.92203 & 20.5368 & 0.19900 & 0.2394 & 0.0250 & -- \\
477 & 13 & 1.13313 & 20.2364 & 0.19900 & 0.2033 & 0.0250 & -- \\
478 & 13 & 0.59300 & 18.9106 & 0.50950 & 0.0880 & 0.0243 & 0.0881\tablenotemark{b} \\ 
485 & 13 & 0.73114 & 19.6791 & 0.19952 & 0.1517 & 0.0245 & -- \\
497 & 13 & 0.53389 & 18.8471 & 0.33775 & 0.0716 & 0.0243 & -- \\
498 & 13 & 0.54786 & 18.3495 & 0.40129 & 0.0655 & 0.0243 & -- \\
501 & 13 & 0.53564 & 19.8623 & 0.20177 & 0.0952 & 0.0243 & 0.1517 \\
502 & 13 & 0.47876 & 17.9946 & 0.16282 & 0.0465 & 0.0244 & -- \\
504 & 13 & 0.53734 & 19.0054 & 0.13891 & 0.0757 & 0.0243 & -- \\
508 & 13 & 0.61470 & 19.6654 & 0.13706 & 0.1151 & 0.0243 & 0.1479 \\
509 & 13 & 0.68490 & 20.2419 & 0.14506 & 0.1609 & 0.0245 & 0.0836 \\
515 & 13 & 0.44215 & 18.3532 & 0.09844 & 0.0443 & 0.0244 & -- \\
520 & 13 & 0.71559 & 19.9856 & 0.04703 & 0.1601 & 0.0245 & 0.1990 \\
523 & 13 & 0.60029 & 18.8808 & 0.14601 & 0.0891 & 0.0243 & 0.1000 \\
525 & 13 & 0.67672 & 19.8333 & 0.13798 & 0.1412 & 0.0243 & -- \\
526 & 13 & 0.54725 & 19.4433 & 0.09844 & 0.0882 & 0.0243 & 0.0835 \\
529 & 13 & 0.52781 & 19.1887 & 0.08734 & 0.0770 & 0.0243 & -- \\
530 & 13 & 0.71692 & 18.6154 & 0.07989 & 0.1103 & 0.0243 & -- \\
532 & 13 & 0.66377 & 19.3781 & 0.21411 & 0.1209 & 0.0243 & -- \\
537 & 13 & 0.83142 & 20.5565 & 0.13948 & 0.2244 & 0.0250 & -- \\
539 & 13 & 0.31046 & 18.2052 & 0.16181 & 0.0232 & 0.0244 & 0.0284 \\
541 & 13 & 1.44808 & 17.8569 & 0.19176 & 0.0496 & 0.0244 & -- \\
546 & 13 & 0.83132 & 19.8455 & 0.19558 & 0.1847 & 0.0245 & -- \\
549 & 13 & 0.99926 & 20.4647 & 0.14110 & 0.2371 & 0.0250 & -- \\
553 & 13 & 0.46492 & 19.2178 & 0.15635 & 0.0615 & 0.0243 & 0.0664 \\
554 & 13 & 0.54244 & 19.0158 & 0.10251 & 0.0772 & 0.0243 & -- \\
556 & 13 & 0.65611 & 18.8411 & 0.09980 & 0.1024 & 0.0243 & -- \\
557 & 13 & 1.37318 & 19.9090 & 0.10900 & 0.1117 & 0.0243 & -- \\
558 & 13 & 0.17091 & 18.2115 & 0.10936 & 0.0846 & 0.0243 & -- \\
558 & 13 & 0.57205 & 19.0006 & 0.10936 & 0.0105 & 0.0244 & -- \\
567 & 13 & 0.59490 & 19.8080 & 0.11074 & 0.1131 & 0.0243 & -- \\
567 & 13 & 0.60708 & 20.1740 & 0.11074 & 0.1295 & 0.0243 & -- \\
569 & 13 & 0.61141 & 19.0390 & 0.07308 & 0.0961 & 0.0243 & 0.0201 \\
574 & 13 & 0.96262 & 20.3645 & 0.03163 & 0.2309 & 0.0250 & 0.1740 \\
579 & 16 & 0.60821 & 19.5383 & 0.06854 & 0.1311 & 0.0289 & -- \\
580 & 13 & 0.47746 & 19.7908 & 0.07691 & 0.0756 & 0.0243 & -- \\
583 & 16 & 0.66862 & 20.0568 & 0.06291 & 0.1651 & 0.0292 & -- \\
586 & 16 & 0.67197 & 19.9786 & 0.05758 & 0.1642 & 0.0292 & 0.1710 \\
587 & 13 & 0.55457 & 19.9488 & 0.06260 & 0.1038 & 0.0243 & 0.1680 \\
590 & 13 & 0.53404 & 20.0988 & 0.05466 & 0.1010 & 0.0243 & -- \\
611 & 13 & 0.61826 & 20.3126 & 0.04781 & 0.1388 & 0.0243 & 0.2880 \\
630 & 13 & 0.69993 & 20.6623 & 0.04882 & 0.1865 & 0.0245 & -- \\
635 & 13 & 0.46777 & 19.3110 & 0.04504 & 0.0638 & 0.0243 & -- \\
647 & 13 & 0.86745 & 20.5141 & 0.02353 & 0.2299 & 0.0250 & -- \\
657 & 13 & 0.06363 & 19.7952 & 0.02951 & 0.0080 & 0.0244 & -- \\
657 & 13 & 0.59567 & 19.3910 & 0.02951 & 0.1011 & 0.0243 & -- \\
658 & 13 & 0.57290 & 19.8568 & 0.03135 & 0.1073 & 0.0243 & 0.0917\tablenotemark{b} \\ 
659 & 13 & 0.62426 & 19.4196 & 0.05547 & 0.1105 & 0.0243 & -- \\
660 & 13 & 0.68376 & 20.5299 & 0.04733 & 0.1736 & 0.0245 & -- \\
665 & 13 & 0.36927 & 19.8699 & 0.04360 & 0.0487 & 0.0244 & 0.1819 \\
668 & 13 & 0.55004 & 19.6021 & 0.05564 & 0.1621 & 0.0245 & 0.1588\tablenotemark{b} \\ 
677 & 13 & 0.70330 & 21.1855 & 0.04541 & 0.2430 & 0.0250 & -- \\
677 & 13 & 1.10518 & 20.7766 & 0.04360 & 0.2168 & 0.0250 & -- \\
687 & 13 & 0.94758 & 20.5253 & 0.02483 & 0.2407 & 0.0250 & -- \\
688 & 13 & 0.66148 & 19.9358 & 0.02399 & 0.1401 & 0.0243 & -- \\
696 & 13 & 0.66665 & 20.2914 & 0.02464 & 0.1563 & 0.0245 & -- \\
699 & 13 & 0.50383 & 19.2307 & 0.05015 & 0.0716 & 0.0243 & 0.0851 \\
706 & 13 & 0.68127 & 20.7678 & 0.05126 & 0.1843 & 0.0245 & -- \\
710 & 13 & 0.82540 & 20.8878 & 0.03391 & 0.2441 & 0.0250 & -- \\
715 & 13 & 0.70071 & 20.9068 & 0.03819 & 0.1998 & 0.0245 & 0.1685 \\
720 & 13 & 1.03609 & 20.8913 & 0.03034 & 0.2635 & 0.0306 & 0.1329\tablenotemark{a} \\ 2
724 & 16 & 0.53159 & 18.5909 & 0.02975 & 0.0895 & 0.0289 & 0.0933\tablenotemark{b} \\ 
732 & 13 & 0.64268 & 20.5079 & 0.04848 & 0.1637 & 0.0245 & 0.2030 \\
732 & 13 & 0.68154 & 20.3330 & 0.04848 & 0.1563 & 0.0245 & 0.2030 \\
734 & 13 & 0.81491 & 20.6592 & 0.03223 & 0.2264 & 0.0250 & 0.0719 \\
741 & 16 & 0.87969 & 19.7788 & 0.02001 & 0.2102 & 0.0287 & -- \\
749 & 13 & 0.69543 & 20.0636 & 0.04477 & 0.1568 & 0.0245 & -- \\
750 & 13 & 0.69561 & 20.2390 & 0.04248 & 0.1583 & 0.0245 & 0.1800 \\
750 & 13 & 0.70324 & 20.0379 & 0.04248 & 0.1646 & 0.0245 & 0.1800 \\
752 & 13 & 0.63009 & 20.6341 & 0.02153 & 0.1565 & 0.0245 & -- \\
752 & 13 & 0.91807 & 20.1178 & 0.02630 & 0.2131 & 0.0250 & -- \\
759 & 16 & 0.94435 & 20.0263 & 0.01980 & 0.2228 & 0.0287 & -- \\
779 & 13 & 0.95300 & 20.9636 & 0.01599 & 0.2717 & 0.0306 & 0.0229\tablenotemark{a} \\ 
781 & 13 & 1.02474 & 20.9976 & 0.02206 & 0.2725 & 0.0306 & 0.2980 \\
791 & 13 & 0.79105 & 20.1921 & 0.03118 & 0.1929 & 0.0245 & -- \\
795 & 13 & 0.54773 & 20.0723 & 0.02871 & 0.1050 & 0.0243 & 0.1359 \\
795 & 13 & 0.62135 & 19.9541 & 0.02962 & 0.1269 & 0.0243 & 0.1359 \\
815 & 13 & 0.90653 & 20.5706 & 0.01982 & 0.2399 & 0.0250 & -- \\
835 & 13 & 0.62725 & 20.1492 & 0.03356 & 0.1360 & 0.0243 & -- \\
850 & 13 & 0.61231 & 20.1526 & 0.03610 & 0.1307 & 0.0243 & -- \\
851 & 16 & 1.26743 & 20.9694 & 0.01711 & 0.1746 & 0.0292 & 0.4069 \\
891 & 13 & 0.75560 & 20.4164 & 0.01970 & 0.1940 & 0.0245 & -- \\
898 & 16 & 0.90464 & 20.6850 & 0.01030 & 0.2464 & 0.0287 & -- \\
937 & 13 & 0.56146 & 19.9056 & 0.04112 & 0.1048 & 0.0243 & -- \\
942 & 13 & 0.58843 & 20.1755 & 0.01418 & 0.1227 & 0.0243 & -- \\
954 & 16 & 0.59112 & 19.3193 & 0.03458 & 0.1207 & 0.0289 & 0.0932 \\
957 & 16 & 0.87561 & 19.4017 & 0.03351 & 0.1971 & 0.0292 & 0.0450\tablenotemark{a} \\ 2
967 & 16 & 0.84000 & 20.1400 & 0.01359 & 0.2169 & 0.0287 & -- \\
969 & 16 & 0.50442 & 18.8411 & 0.03372 & 0.0852 & 0.0289 & -- \\
986 & 13 & 0.85127 & 20.6757 & 0.04148 & 0.2368 & 0.0250 & -- \\
1015 & 13 & 1.29165 & 20.5934 & 0.01384 & 0.1684 & 0.0245 & -- \\
1019 & 16 & 0.72710 & 20.3242 & 0.01938 & 0.1934 & 0.0292 & -- \\
1030 & 16 & 0.85723 & 20.3117 & 0.01841 & 0.2261 & 0.0287 & 0.1780\tablenotemark{b} \\ 
1034 & 13 & 0.72808 & 20.0985 & 0.03329 & 0.1692 & 0.0245 & -- \\
1050 & 16 & 0.58565 & 19.3148 & 0.01948 & 0.1188 & 0.0289 & 0.1208\tablenotemark{b} \\ 
1062 & 13 & 0.76750 & 20.5136 & 0.03201 & 0.2033 & 0.0250 & -- \\
1063 & 13 & 0.68803 & 20.2837 & 0.04027 & 0.1639 & 0.0245 & -- \\
1081 & 13 & 0.57394 & 20.0114 & 0.01671 & 0.1123 & 0.0243 & 0.1585 \\
1095 & 13 & 0.69758 & 20.4506 & 0.02989 & 0.1752 & 0.0245 & 0.2108\tablenotemark{b} \\ 
1095 & 13 & 0.73707 & 20.2775 & 0.02989 & 0.1808 & 0.0245 & 0.2108\tablenotemark{b} \\ 
1101 & 16 & 0.92782 & 20.3118 & 0.01740 & 0.2331 & 0.0287 & 0.2322 \\
1114 & 13 & 0.47207 & 20.1248 & 0.02872 & 0.0811 & 0.0243 & 0.0140 \\
1120 & 13 & 0.79550 & 20.9401 & 0.02972 & 0.2383 & 0.0250 & 0.2218 \\
1140 & 13 & 1.17952 & 20.9944 & 0.01990 & 0.2347 & 0.0250 & -- \\
1299 & 13 & 0.41872 & 20.1729 & 0.02565 & 0.0661 & 0.0243 & 0.2247 \\
1345 & 13 & 0.50007 & 19.9364 & 0.04435 & 0.0856 & 0.0243 & 0.1095 \\
1356 & 13 & 0.44753 & 20.0358 & 0.04992 & 0.0718 & 0.0243 & 0.0698 \\
1413 & 13 & 0.57322 & 19.6151 & 0.02308 & 0.1005 & 0.0243 & 0.1427 \\
1441 & 13 & 0.56523 & 19.5451 & 0.01963 & 0.0961 & 0.0243 & -- \\
1445 & 13 & 0.60697 & 20.4888 & 0.02511 & 0.1411 & 0.0243 & 0.1694 \\
1475 & 13 & 0.71660 & 20.7940 & 0.03053 & 0.2002 & 0.0250 & -- \\
1481 & 13 & 0.46412 & 20.1934 & 0.03169 & 0.0801 & 0.0243 & -- \\
1487 & 13 & 0.80138 & 20.7258 & 0.01770 & 0.2265 & 0.0250 & 0.2111 \\
1489 & 13 & 1.03265 & 19.9782 & 0.02366 & 0.2055 & 0.0250 & 0.2060 \\
1495 & 13 & 0.59984 & 20.0652 & 0.02300 & 0.1231 & 0.0243 & 0.1429 \\
1497 & 13 & 0.65830 & 20.3739 & 0.02513 & 0.1567 & 0.0245 & 0.1669 \\
1499 & 13 & 0.60883 & 20.1277 & 0.03933 & 0.1285 & 0.0243 & 0.1569 \\
1526 & 13 & 0.52537 & 19.6862 & 0.04942 & 0.0875 & 0.0243 & 0.0800 \\
1551 & 16 & 0.70359 & 19.6431 & 0.01295 & 0.1658 & 0.0292 & -- \\
1577 & 13 & 0.59981 & 19.8800 & 0.02463 & 0.1170 & 0.0243 & 0.1409\tablenotemark{b} \\ 
1578 & 16 & 0.63441 & 19.1858 & 0.02452 & 0.1323 & 0.0289 & -- \\
1592 & 16 & 0.86737 & 20.3286 & 0.01626 & 0.2283 & 0.0287 & -- \\
1608 & 16 & 0.73723 & 19.3522 & 0.01494 & 0.1678 & 0.0292 & 0.1319\tablenotemark{b} \\ 
1613 & 16 & 0.67606 & 19.7880 & 0.01833 & 0.1606 & 0.0292 & 0.1608\tablenotemark{b} \\ 
1620 & 13 & 0.45548 & 19.8524 & 0.02533 & 0.0705 & 0.0243 & 0.0821 \\
1647 & 16 & 0.78172 & 19.9377 & 0.02458 & 0.1972 & 0.0292 & -- \\
1656 & 13 & 0.42058 & 18.5126 & 0.00840 & 0.0423 & 0.0244 & 0.0231 \\
1657 & 16 & 0.77811 & 19.6795 & 0.02757 & 0.1881 & 0.0292 & -- \\
1661 & 16 & 0.66630 & 19.6312 & 0.01671 & 0.1532 & 0.0292 & 0.1690 \\
1661 & 16 & 0.68407 & 19.1141 & 0.01671 & 0.1463 & 0.0289 & 0.1690 \\
1670 & 16 & 1.19250 & 19.5283 & 0.02222 & 0.1627 & 0.0292 & -- \\
1677 & 16 & 0.87111 & 19.1111 & 0.00942 & 0.1875 & 0.0292 & 0.1820 \\
1680 & 13 & 0.48350 & 19.8260 & 0.01743 & 0.0781 & 0.0243 & -- \\
1694 & 16 & 0.82213 & 19.9190 & 0.01146 & 0.2058 & 0.0287 & -- \\
1711 & 13 & 0.46982 & 19.7342 & 0.02744 & 0.0723 & 0.0243 & -- \\
1712 & 16 & 1.01113 & 20.7184 & 0.01231 & 0.2457 & 0.0287 & -- \\
1714 & 16 & 0.96882 & 20.6158 & 0.01342 & 0.2242 & 0.0287 & -- \\
1714 & 16 & 1.09522 & 20.5804 & 0.01342 & 0.2450 & 0.0287 & -- \\
1717 & 13 & 0.64266 & 20.0143 & 0.01437 & 0.1365 & 0.0243 & -- \\
1719 & 13 & 1.01227 & 19.7564 & 0.00958 & 0.1947 & 0.0245 & -- \\
1728 & 13 & 0.44244 & 18.8809 & 0.02792 & 0.0513 & 0.0243 & -- \\
1739 & 16 & 0.56328 & 19.5497 & 0.01693 & 0.1157 & 0.0289 & -- \\
1746 & 13 & 0.74077 & 20.3378 & 0.00950 & 0.1850 & 0.0245 & -- \\
1752 & 13 & 0.53579 & 20.5101 & 0.01115 & 0.1137 & 0.0243 & -- \\
1759 & 16 & 1.09547 & 19.6304 & 0.02935 & 0.1919 & 0.0292 & 0.1680 \\
1760 & 16 & 0.61888 & 19.5574 & 0.02926 & 0.1352 & 0.0289 & 0.1711 \\
1785 & 13 & 0.60443 & 20.1930 & 0.01169 & 0.1292 & 0.0243 & 0.2136 \\
1795 & 16 & 0.57284 & 19.9279 & 0.01240 & 0.1266 & 0.0289 & 0.0631 \\
1799 & 13 & 0.91528 & 20.7716 & 0.01594 & 0.2546 & 0.0306 & 0.2451 \\
1810 & 16 & 0.68330 & 19.5209 & 0.01207 & 0.1561 & 0.0292 & -- \\
1813 & 16 & 0.54633 & 19.0667 & 0.01668 & 0.1015 & 0.0289 & 0.0947\tablenotemark{b} \\ 
1820 & 13 & 0.81117 & 20.6951 & 0.02743 & 0.2275 & 0.0250 & -- \\
1821 & 16 & 0.61573 & 19.3145 & 0.01523 & 0.1289 & 0.0289 & -- \\
1826 & 16 & 0.86756 & 19.8284 & 0.01636 & 0.2104 & 0.0287 & -- \\
1856 & 13 & 0.56361 & 20.5709 & 0.01695 & 0.1267 & 0.0243 & 0.1854\tablenotemark{b} \\ 
1874 & 16 & 0.73527 & 19.2227 & 0.01704 & 0.1638 & 0.0292 & -- \\
1889 & 16 & 0.43053 & 19.5939 & 0.01446 & 0.0730 & 0.0289 & 0.1860\tablenotemark{b} \\
1900 & 16 & 0.73368 & 19.3106 & 0.01198 & 0.1657 & 0.0292 & 0.1718\tablenotemark{b} \\ 
1905 & 13 & 0.90926 & 20.5457 & 0.01967 & 0.2386 & 0.0250 & 0.3392\tablenotemark{b} \\ 
1909 & 13 & 0.26985 & 18.7493 & 0.02100 & 0.0217 & 0.0244 & 0.1456\tablenotemark{a} \\ 
1914 & 13 & 0.70911 & 19.9440 & 0.02253 & 0.1562 & 0.0245 & 0.1712 \\
1917 & 13 & 0.65653 & 20.7987 & 0.03411 & 0.1753 & 0.0245 & -- \\
1926 & 13 & 0.45225 & 19.0446 & 0.03235 & 0.0558 & 0.0243 & 0.1338 \\
1934 & 13 & 0.81717 & 20.5205 & 0.01296 & 0.2185 & 0.0250 & 0.2194 \\
1954 & 13 & 0.49161 & 20.5259 & 0.01765 & 0.0976 & 0.0243 & 0.1810 \\
1965 & 16 & 1.33724 & 20.8859 & 0.01321 & 0.1213 & 0.0289 & -- \\
1965 & 16 & 1.39209 & 21.0136 & 0.01321 & 0.1420 & 0.0289 & -- \\
1965 & 16 & 1.52184 & 20.8693 & 0.01321 & 0.0707 & 0.0289 & -- \\
1979 & 13 & 0.69590 & 19.8762 & 0.01644 & 0.1491 & 0.0243 & 0.1687 \\
1987 & 16 & 0.60901 & 19.0374 & 0.01949 & 0.1210 & 0.0289 & -- \\
1990 & 13 & 0.83691 & 20.4901 & 0.02416 & 0.2217 & 0.0250 & 0.1269 \\
2005 & 16 & 0.65914 & 19.1451 & 0.02643 & 0.1393 & 0.0289 & 0.1234 \\
2008 & 13 & 0.75044 & 20.3750 & 0.05147 & 0.1902 & 0.0245 & 0.1810 \\
2016 & 13 & 0.65029 & 20.2297 & 0.04019 & 0.1477 & 0.0243 & -- \\
2017 & 13 & 0.62659 & 20.1667 & 0.04933 & 0.1364 & 0.0243 & 0.1187 \\
2034 & 16 & 0.68979 & 18.7973 & 0.01520 & 0.1405 & 0.0289 & 0.1130\tablenotemark{b} \\ 
2042 & 16 & 0.59510 & 19.7934 & 0.01658 & 0.1319 & 0.0289 & 0.2353\tablenotemark{a} \\ 
2063 & 13 & 0.66303 & 19.9529 & 0.03458 & 0.1412 & 0.0243 & 0.0353\tablenotemark{a} \\
2065 & 16 & 0.39157 & 19.3454 & 0.04122 & 0.0595 & 0.0289 & 0.0726 \\
2069 & 16 & 0.50431 & 19.3059 & 0.02346 & 0.0918 & 0.0289 & 0.1160 \\
2110 & 16 & 0.51463 & 19.0823 & 0.02508 & 0.0917 & 0.0289 & 0.0980 \\
2111 & 16 & 0.97285 & 20.2236 & 0.02595 & 0.2296 & 0.0287 & 0.2290 \\
2116 & 13 & 1.29418 & 20.4549 & 0.03000 & 0.1611 & 0.0245 & -- \\
2126 & 13 & 0.71284 & 20.4204 & 0.06225 & 0.1793 & 0.0245 & 0.1656\tablenotemark{b} \\ 
2129 & 13 & 0.74726 & 20.3378 & 0.05130 & 0.1872 & 0.0245 & -- \\
2141 & 16 & 0.60366 & 20.0573 & 0.02138 & 0.1410 & 0.0289 & 0.1584\tablenotemark{b} \\ 
2141 & 16 & 0.79995 & 19.6468 & 0.02125 & 0.1923 & 0.0292 & 0.1584\tablenotemark{b} \\ 
2147 & 13 & 0.48392 & 19.8084 & 0.03332 & 0.0727 & 0.0243 & 0.0350 \\
2147 & 13 & 0.48680 & 19.5180 & 0.03332 & 0.0779 & 0.0243 & 0.0350 \\
2152 & 13 & 0.41326 & 19.6833 & 0.04018 & 0.0565 & 0.0243 & 0.0410 \\
2155 & 13 & 1.17570 & 20.4862 & 0.06581 & 0.2054 & 0.0250 & 0.2465\tablenotemark{b} \\ 
2157 & 13 & 1.02867 & 21.0270 & 0.01809 & 0.2743 & 0.0306 & -- \\
2160 & 16 & 0.80138 & 20.2053 & 0.03127 & 0.2110 & 0.0287 & -- \\
2162 & 13 & 0.43867 & 20.1324 & 0.03728 & 0.0711 & 0.0243 & 0.0322 \\
2170 & 13 & 0.49221 & 19.2521 & 0.08171 & 0.0690 & 0.0243 & 0.1030 \\
2173 & 13 & 0.61042 & 20.2919 & 0.06308 & 0.1350 & 0.0243 & -- \\
2177 & 13 & 0.67667 & 20.4227 & 0.03891 & 0.1659 & 0.0245 & 0.1610 \\
2178 & 16 & 0.53237 & 19.8593 & 0.05602 & 0.1105 & 0.0289 & 0.0928 \\
2188 & 16 & 0.47539 & 20.1565 & 0.02184 & 0.0952 & 0.0289 & -- \\
2192 & 16 & 0.61924 & 20.2626 & 0.01013 & 0.1519 & 0.0292 & 0.1875 \\
2193 & 13 & 0.45168 & 20.0398 & 0.04102 & 0.0731 & 0.0243 & -- \\
2195 & 13 & 0.86845 & 20.3751 & 0.01547 & 0.2215 & 0.0250 & -- \\
2196 & 16 & 0.52552 & 19.8127 & 0.00573 & 0.1072 & 0.0289 & 0.1339 \\
2197 & 13 & 0.46110 & 18.2195 & 0.00777 & 0.0461 & 0.0244 & 0.0308 \\
2198 & 13 & 0.90143 & 20.8216 & 0.00694 & 0.2562 & 0.0306 & 0.0798 \\
2199 & 13 & 0.61377 & 19.3977 & 0.01070 & 0.1067 & 0.0243 & 0.0299 \\
2199 & 16 & 0.56860 & 19.0855 & 0.01070 & 0.1089 & 0.0289 & 0.0299 \\
2200 & 13 & 0.70059 & 20.0345 & 0.06997 & 0.1573 & 0.0245 & -- \\
2205 & 13 & 0.52694 & 19.8040 & 0.05611 & 0.0909 & 0.0243 & 0.0876 \\
2215 & 13 & 0.83647 & 20.6864 & 0.02702 & 0.2339 & 0.0250 & -- \\
2217 & 16 & 0.67713 & 19.6565 & 0.05380 & 0.1575 & 0.0292 & -- \\
2219 & 13 & 0.97399 & 20.2862 & 0.02349 & 0.2262 & 0.0250 & 0.2256 \\
2221 & 13 & 0.75515 & 20.2772 & 0.01248 & 0.1866 & 0.0245 & 0.1019\tablenotemark{b} \\ 
2228 & 13 & 0.62752 & 19.6874 & 0.04829 & 0.1200 & 0.0243 & 0.1013 \\
2229 & 13 & 0.75386 & 20.3148 & 0.02118 & 0.1882 & 0.0245 & -- \\
2238 & 13 & 0.67963 & 20.1779 & 0.01715 & 0.1562 & 0.0245 & -- \\
2240 & 13 & 0.74809 & 20.7689 & 0.04179 & 0.2110 & 0.0250 & 0.1380 \\
2241 & 16 & 0.51370 & 19.5172 & 0.03004 & 0.0982 & 0.0289 & -- \\
2243 & 13 & 0.63968 & 20.1087 & 0.02133 & 0.1390 & 0.0243 & -- \\
2244 & 13 & 0.57937 & 19.8780 & 0.02313 & 0.1100 & 0.0243 & 0.0968 \\
2246 & 13 & 1.03852 & 20.8495 & 0.02377 & 0.2602 & 0.0306 & 0.2250 \\
2251 & 16 & 0.72084 & 20.1940 & 0.04985 & 0.1873 & 0.0292 & -- \\
2252 & 13 & 0.79111 & 20.6880 & 0.01414 & 0.2210 & 0.0250 & 0.1147 \\
2254 & 13 & 0.70617 & 20.2366 & 0.05330 & 0.1682 & 0.0245 & 0.1780 \\
2255 & 13 & 0.68960 & 19.6058 & 0.02622 & 0.1366 & 0.0243 & 0.0806 \\
2256 & 13 & 0.53847 & 19.2448 & 0.04982 & 0.0811 & 0.0243 & 0.0581 \\
2257 & 13 & 0.62971 & 19.4428 & 0.04085 & 0.1128 & 0.0243 & 0.1054 \\
2261 & 16 & 0.56827 & 20.2002 & 0.04453 & 0.1306 & 0.0289 & 0.2240\tablenotemark{b} \\ 
2262 & 13 & 0.91988 & 20.5466 & 0.05507 & 0.2399 & 0.0250 & -- \\
2262 & 16 & 0.89261 & 20.1596 & 0.05507 & 0.2251 & 0.0287 & -- \\
2263 & 13 & 0.60642 & 20.2462 & 0.04994 & 0.1319 & 0.0243 & 0.1051 \\
2266 & 16 & 0.71157 & 20.3038 & 0.04183 & 0.1875 & 0.0292 & 0.1671 \\
2267 & 13 & 0.86400 & 20.5473 & 0.02297 & 0.2313 & 0.0250 & -- \\
2268 & 13 & 0.73229 & 20.4696 & 0.03480 & 0.1888 & 0.0245 & -- \\
2269 & 13 & 1.16571 & 21.0218 & 0.03386 & 0.2415 & 0.0250 & -- \\
2270 & 16 & 0.78089 & 20.7234 & 0.03440 & 0.2240 & 0.0287 & 0.2377 \\
2272 & 13 & 0.62073 & 20.2334 & 0.05035 & 0.1368 & 0.0243 & 0.1329 \\
2273 & 13 & 1.01048 & 20.7064 & 0.03138 & 0.2527 & 0.0306 & -- \\
2274 & 13 & 1.02068 & 19.8009 & 0.03767 & 0.1966 & 0.0245 & -- \\
2275 & 13 & 0.52463 & 19.7815 & 0.04312 & 0.0896 & 0.0243 & 0.1029 \\
2278 & 13 & 0.73838 & 20.6429 & 0.03667 & 0.2003 & 0.0250 & -- \\
2279 & 13 & 0.65266 & 19.8254 & 0.07495 & 0.1330 & 0.0243 & -- \\
2285 & 16 & 0.43095 & 19.1077 & 0.02784 & 0.0675 & 0.0289 & -- \\
2286 & 13 & 0.53359 & 19.6268 & 0.04234 & 0.0886 & 0.0243 & -- \\
2288 & 13 & 0.48939 & 19.3434 & 0.06314 & 0.0700 & 0.0243 & -- \\
2289 & 13 & 0.88125 & 20.6091 & 0.04916 & 0.2385 & 0.0250 & 0.2276 \\
2291 & 13 & 0.81070 & 19.9702 & 0.05685 & 0.1864 & 0.0245 & 0.1810 \\
2292 & 13 & 0.64847 & 18.9927 & 0.04306 & 0.1047 & 0.0243 & 0.1190\tablenotemark{b} \\ 
2297 & 13 & 0.50607 & 20.0410 & 0.03817 & 0.0901 & 0.0243 & -- \\
2297 & 13 & 0.74219 & 20.3259 & 0.04254 & 0.1849 & 0.0245 & -- \\
2298 & 13 & 0.80612 & 20.0823 & 0.05148 & 0.1911 & 0.0245 & -- \\
2299 & 13 & 0.55541 & 20.0556 & 0.03908 & 0.1515 & 0.0245 & -- \\
2299 & 13 & 0.69345 & 19.9546 & 0.05563 & 0.1071 & 0.0243 & -- \\
2300 & 13 & 0.64421 & 19.5768 & 0.07032 & 0.1216 & 0.0243 & -- \\
2310 & 13 & 0.63194 & 19.3633 & 0.07897 & 0.1111 & 0.0243 & -- \\
2311 & 13 & 0.52620 & 18.9747 & 0.06498 & 0.0722 & 0.0243 & 0.0890 \\
2314 & 13 & 0.86960 & 20.0638 & 0.07862 & 0.2035 & 0.0250 & -- \\
2315 & 13 & 0.52220 & 19.0739 & 0.08541 & 0.0732 & 0.0243 & 0.0894 \\
2316 & 13 & 0.99357 & 20.4234 & 0.09037 & 0.2346 & 0.0250 & 0.2147 \\
2317 & 13 & 0.89940 & 20.4905 & 0.06625 & 0.2337 & 0.0250 & 0.2110 \\
2318 & 13 & 0.60557 & 20.1790 & 0.06819 & 0.1291 & 0.0243 & 0.1405 \\
2319 & 13 & 0.54066 & 19.3494 & 0.11363 & 0.0841 & 0.0243 & 0.0557\tablenotemark{b} \\ 
2320 & 13 & 0.62367 & 19.0166 & 0.17918 & 0.0988 & 0.0243 & 0.1710 \\
2321 & 13 & 0.66613 & 20.2440 & 0.16135 & 0.1541 & 0.0245 & -- \\
2322 & 13 & 0.69446 & 19.7792 & 0.25078 & 0.1447 & 0.0243 & -- \\
2323 & 13 & 0.66152 & 19.7576 & 0.09939 & 0.1334 & 0.0243 & -- \\
2326 & 13 & 0.65102 & 19.4072 & 0.48845 & 0.1181 & 0.0243 & -- \\
2327 & 13 & 0.61473 & 19.8558 & 0.08081 & 0.1213 & 0.0243 & -- \\
2349 & 16 & 0.38086 & 19.0358 & 0.04782 & 0.0540 & 0.0289 & -- \\
2349 & 16 & 0.58182 & 20.0422 & 0.04858 & 0.1324 & 0.0289 & -- \\
2353 & 13 & 0.52112 & 19.2298 & 0.05477 & 0.0761 & 0.0243 & 0.1210 \\
2355 & 13 & 1.04350 & 20.6513 & 0.06633 & 0.2458 & 0.0250 & 0.1244 \\
2355 & 16 & 0.59421 & 20.4228 & 0.06633 & 0.1459 & 0.0289 & 0.1244 \\
2356 & 16 & 0.54219 & 19.3538 & 0.06931 & 0.1050 & 0.0289 & 0.1161 \\
2359 & 13 & 0.73060 & 20.3196 & 0.08706 & 0.1807 & 0.0245 & -- \\
2373 & 13 & 0.80311 & 20.8274 & 0.05509 & 0.2335 & 0.0250 & -- \\
2379 & 13 & 0.68978 & 19.8457 & 0.11891 & 0.1459 & 0.0243 & -- \\
2381 & 13 & 0.83569 & 20.2282 & 0.05782 & 0.2061 & 0.0250 & 0.0726\tablenotemark{a} \\ 
2386 & 13 & 0.56365 & 19.9434 & 0.07340 & 0.1067 & 0.0243 & -- \\
2386 & 13 & 0.73147 & 20.4833 & 0.06705 & 0.1893 & 0.0245 & -- \\
2387 & 13 & 0.80581 & 19.0725 & 0.14370 & 0.1449 & 0.0243 & 0.1450 \\
2388 & 13 & 0.46280 & 17.7167 & 0.05006 & 0.0405 & 0.0244 & 0.0615 \\
2390 & 16 & 0.74397 & 20.4617 & 0.11311 & 0.2034 & 0.0287 & 0.2280 \\
2390 & 16 & 0.81282 & 20.4291 & 0.11562 & 0.2216 & 0.0287 & 0.2280 \\
2392 & 16 & 0.83911 & 20.1990 & 0.07231 & 0.2189 & 0.0287 & -- \\
2395 & 13 & 0.72047 & 20.0539 & 0.04946 & 0.1647 & 0.0245 & 0.1508\tablenotemark{b} \\ 
2396 & 13 & 0.56837 & 19.7775 & 0.07442 & 0.1035 & 0.0243 & 0.1946 \\
2397 & 13 & 0.81644 & 20.4615 & 0.05322 & 0.2149 & 0.0250 & 0.2240 \\
2397 & 16 & 0.91703 & 20.0800 & 0.05322 & 0.2240 & 0.0287 & 0.2240 \\
2406 & 16 & 0.79625 & 20.4579 & 0.06294 & 0.2186 & 0.0287 & -- \\
2407 & 13 & 0.62568 & 19.5082 & 0.05697 & 0.1136 & 0.0243 & -- \\
2407 & 13 & 0.65888 & 19.8073 & 0.05697 & 0.1344 & 0.0243 & -- \\
2408 & 13 & 0.72784 & 20.4625 & 0.05279 & 0.1869 & 0.0245 & -- \\
2409 & 13 & 0.67936 & 20.2982 & 0.10835 & 0.1613 & 0.0245 & 0.1479 \\
2413 & 13 & 0.64357 & 20.3016 & 0.08163 & 0.1480 & 0.0243 & -- \\
2414 & 13 & 0.54429 & 19.6266 & 0.05859 & 0.1143 & 0.0243 & -- \\
2414 & 13 & 0.66671 & 19.1478 & 0.05859 & 0.0918 & 0.0243 & -- \\
2419 & 13 & 0.65240 & 20.2949 & 0.05203 & 0.1511 & 0.0245 & 0.0456 \\
2422 & 16 & 0.98668 & 20.5918 & 0.12078 & 0.2430 & 0.0287 & -- \\
2423 & 16 & 0.53334 & 19.5092 & 0.08588 & 0.1046 & 0.0289 & -- \\
2424 & 13 & 0.74243 & 20.2355 & 0.06700 & 0.1804 & 0.0245 & 0.1510 \\
2425 & 13 & 0.81797 & 20.7676 & 0.09094 & 0.2341 & 0.0250 & -- \\
2425 & 16 & 1.13616 & 20.2419 & 0.09307 & 0.2011 & 0.0287 & -- \\
2429 & 16 & 0.78587 & 20.3804 & 0.06104 & 0.2131 & 0.0287 & -- \\
2429 & 16 & 1.10455 & 20.6698 & 0.06104 & 0.2250 & 0.0287 & -- \\
2431 & 16 & 0.63601 & 20.2324 & 0.06324 & 0.1576 & 0.0292 & -- \\
2431 & 16 & 0.70041 & 20.0241 & 0.06324 & 0.1754 & 0.0292 & -- \\
2432 & 13 & 0.68740 & 20.2438 & 0.09827 & 0.1619 & 0.0245 & -- \\
2435 & 16 & 0.73958 & 20.1356 & 0.06413 & 0.1915 & 0.0292 & -- \\
2437 & 13 & 0.69214 & 20.2262 & 0.08662 & 0.1628 & 0.0245 & -- \\
2439 & 16 & 0.70738 & 20.2025 & 0.06611 & 0.1830 & 0.0292 & -- \\
2440 & 16 & 0.57093 & 19.7181 & 0.08757 & 0.1216 & 0.0289 & 0.0906 \\
2443 & 13 & 0.64240 & 19.8860 & 0.06173 & 0.1317 & 0.0243 & 0.1080 \\
2445 & 13 & 0.64582 & 20.3701 & 0.05318 & 0.1517 & 0.0245 & -- \\
2447 & 13 & 0.44196 & 19.2650 & 0.09851 & 0.0568 & 0.0243 & -- \\
2447 & 13 & 0.62960 & 20.3519 & 0.09851 & 0.1447 & 0.0243 & -- \\
2449 & 13 & 0.49500 & 19.6080 & 0.05987 & 0.1072 & 0.0243 & -- \\
2449 & 13 & 0.58114 & 19.7637 & 0.05987 & 0.0768 & 0.0243 & -- \\
2454 & 13 & 0.65046 & 20.2845 & 0.13473 & 0.1500 & 0.0243 & 0.1590 \\
2454 & 16 & 0.61853 & 19.8051 & 0.13443 & 0.1407 & 0.0289 & 0.1590 \\
2454 & 16 & 0.65777 & 19.9817 & 0.13443 & 0.1592 & 0.0292 & 0.1590 \\
2457 & 16 & 0.50568 & 19.0917 & 0.08380 & 0.0891 & 0.0289 & 0.0597 \\
2458 & 16 & 0.55545 & 19.8929 & 0.04334 & 0.1195 & 0.0289 & -- \\
2471 & 13 & 0.66070 & 20.0663 & 0.10078 & 0.1449 & 0.0243 & 0.1078 \\
2471 & 16 & 0.78578 & 19.8804 & 0.10078 & 0.1964 & 0.0292 & 0.1078 \\
2472 & 13 & 1.08891 & 21.0285 & 0.04027 & 0.2643 & 0.0306 & -- \\
2475 & 13 & 0.82174 & 20.5048 & 0.11083 & 0.2188 & 0.0250 & -- \\
2483 & 16 & 0.52421 & 19.9429 & 0.06316 & 0.1090 & 0.0289 & -- \\
2491 & 16 & 0.69385 & 20.2091 & 0.06395 & 0.1784 & 0.0292 & -- \\
2494 & 13 & 0.68420 & 19.4728 & 0.11406 & 0.1301 & 0.0243 & -- \\
2495 & 13 & 0.52068 & 19.2362 & 0.07727 & 0.1030 & 0.0243 & 0.0775 \\
2495 & 13 & 0.57994 & 19.6307 & 0.07727 & 0.0761 & 0.0243 & 0.0775 \\
2503 & 13 & 0.44194 & 19.5523 & 0.06396 & 0.0615 & 0.0243 & 0.0827 \\
2505 & 13 & 0.44775 & 20.4853 & 0.06581 & 0.0813 & 0.0243 & -- \\
2506 & 13 & 0.65002 & 20.5516 & 0.05708 & 0.1612 & 0.0245 & 0.0289 \\
2507 & 13 & 0.76623 & 20.1576 & 0.09621 & 0.1840 & 0.0245 & 0.1960 \\
2512 & 13 & 0.58753 & 20.1112 & 0.08314 & 0.1202 & 0.0243 & 0.1603 \\
2513 & 13 & 0.74724 & 20.5557 & 0.09173 & 0.1987 & 0.0245 & 0.0250 \\
2515 & 13 & 0.84587 & 20.0317 & 0.07771 & 0.1974 & 0.0245 & -- \\
2516 & 13 & 0.64835 & 20.2746 & 0.10313 & 0.1488 & 0.0243 & 0.0793\tablenotemark{a,b} \\
2516 & 16 & 0.64359 & 20.3556 & 0.10313 & 0.1638 & 0.0292 & 0.0793\tablenotemark{a,b} \\
2517 & 13 & 0.76274 & 19.5212 & 0.11367 & 0.1537 & 0.0245 & -- \\
2522 & 13 & 0.66416 & 20.4527 & 0.09737 & 0.1624 & 0.0245 & 0.1562 \\
2530 & 13 & 0.56802 & 19.7513 & 0.11501 & 0.1026 & 0.0243 & -- \\
2532 & 13 & 0.69960 & 20.4157 & 0.08593 & 0.1742 & 0.0245 & -- \\
2535 & 13 & 0.66319 & 19.5869 & 0.14319 & 0.1278 & 0.0243 & -- \\
2545 & 13 & 0.76899 & 19.8956 & 0.07205 & 0.1720 & 0.0245 & -- \\
2551 & 13 & 0.50094 & 20.1173 & 0.07791 & 0.0903 & 0.0243 & -- \\
2552 & 16 & 0.68599 & 20.2365 & 0.05088 & 0.1764 & 0.0292 & 0.1330\tablenotemark{b} \\ 
2562 & 13 & 0.80266 & 19.7829 & 0.06523 & 0.2218 & 0.0250 & -- \\
2562 & 13 & 0.90301 & 20.2926 & 0.06446 & 0.1753 & 0.0245 & -- \\
2570 & 13 & 0.53636 & 18.9403 & 0.04200 & 0.0741 & 0.0243 & -- \\
2571 & 16 & 0.54843 & 19.5510 & 0.04538 & 0.1105 & 0.0289 & 0.1084\tablenotemark{b} \\ 
2574 & 16 & 0.53587 & 20.3121 & 0.05075 & 0.1203 & 0.0289 & -- \\
2582 & 16 & 0.60813 & 19.8203 & 0.04932 & 0.1373 & 0.0289 & -- \\
2584 & 13 & 0.58608 & 20.3071 & 0.11019 & 0.1263 & 0.0243 & 0.1200 \\
2590 & 16 & 0.68153 & 20.2835 & 0.03588 & 0.1762 & 0.0292 & 0.0790\tablenotemark{a} \\ 
2594 & 13 & 0.64749 & 20.4907 & 0.09116 & 0.1575 & 0.0245 & -- \\
2602 & 13 & 0.66981 & 20.0542 & 0.07331 & 0.1476 & 0.0243 & -- \\
2607 & 13 & 0.59903 & 19.9719 & 0.05932 & 0.1197 & 0.0243 & 0.1201 \\
2610 & 13 & 0.69146 & 20.2495 & 0.04707 & 0.1636 & 0.0245 & -- \\
2616 & 13 & 0.67396 & 20.1020 & 0.06841 & 0.1510 & 0.0245 & 0.1832 \\
2616 & 16 & 0.70984 & 19.9135 & 0.06841 & 0.1753 & 0.0292 & 0.1832 \\
2617 & 16 & 0.68110 & 20.3109 & 0.06647 & 0.1768 & 0.0292 & 0.1630\tablenotemark{b} \\ 
2620 & 13 & 1.14380 & 20.9492 & 0.07324 & 0.2439 & 0.0250 & 0.1911 \\
2621 & 13 & 0.97878 & 20.6458 & 0.06112 & 0.2496 & 0.0250 & -- \\
2622 & 13 & 0.52331 & 19.0075 & 0.05718 & 0.0722 & 0.0243 & 0.0621 \\
2623 & 13 & 0.84819 & 20.3273 & 0.06631 & 0.2146 & 0.0250 & 0.1784 \\
2624 & 13 & 0.89365 & 20.6794 & 0.07054 & 0.2452 & 0.0250 & -- \\
2627 & 13 & 0.58119 & 20.1524 & 0.07086 & 0.1193 & 0.0243 & 0.1255 \\
2631 & 13 & 1.09759 & 20.7569 & 0.03792 & 0.2434 & 0.0250 & 0.2730 \\
2631 & 16 & 0.98491 & 20.6245 & 0.03859 & 0.2445 & 0.0287 & 0.2730 \\
2633 & 16 & 0.60596 & 19.7828 & 0.05176 & 0.1356 & 0.0289 & -- \\
2639 & 16 & 0.74011 & 19.8492 & 0.05066 & 0.1829 & 0.0292 & -- \\
2649 & 13 & 0.50078 & 19.5811 & 0.07831 & 0.0779 & 0.0243 & -- \\
2650 & 13 & 0.73413 & 19.3363 & 0.06955 & 0.1389 & 0.0243 & -- \\
2657 & 13 & 0.46187 & 18.4020 & 0.12911 & 0.0486 & 0.0244 & 0.0402 \\
2666 & 13 & 0.39746 & 18.4351 & 0.03879 & 0.0374 & 0.0244 & 0.0272 \\
2668 & 16 & 0.46489 & 19.5665 & 0.03653 & 0.0831 & 0.0289 & -- \\
2668 & 16 & 0.57262 & 19.6900 & 0.03718 & 0.1217 & 0.0289 & -- \\
2671 & 13 & 0.67403 & 20.5999 & 0.05555 & 0.1731 & 0.0245 & 0.1799 \\
2672 & 13 & 0.66410 & 19.2755 & 0.04222 & 0.1176 & 0.0243 & 0.2412 \\
2675 & 16 & 0.58614 & 19.5086 & 0.07751 & 0.1228 & 0.0289 & 0.0713 \\
2695 & 16 & 0.53040 & 19.3407 & 0.03076 & 0.1008 & 0.0289 & -- \\
2696 & 13 & 0.56125 & 19.3736 & 0.03027 & 0.0906 & 0.0243 & 0.0844 \\
2698 & 16 & 0.64731 & 19.5626 & 0.02219 & 0.1451 & 0.0289 & 0.0979 \\
2706 & 16 & 0.57138 & 19.4338 & 0.08982 & 0.1163 & 0.0289 & -- \\
2711 & 13 & 0.74385 & 20.1832 & 0.04961 & 0.1783 & 0.0245 & -- \\
\enddata
\tablenotetext{a}{The photometric redshift estimate differs significantly from the single-galaxy spectroscopic redshift.}
\tablenotetext{b}{This spectroscopic redshift was obtained after the photometric redshift estimate, and was not used in the derivation.}
\end{deluxetable}

\end{document}